\documentclass[journal,final,twocolumn,10pt]{IEEEtran}
\usepackage[utf8x]{inputenc}
\usepackage[T1]{fontenc}
\usepackage[font=small]{caption}
\usepackage{graphicx}
\usepackage{color}
\usepackage{tabularx}
\usepackage{multirow}
\usepackage{booktabs}
\usepackage{amsmath}
\usepackage[plainpages=false,pdfpagelabels=true,pdftitle={HEAP: Reliable Assessment of BGP Hijacking Attacks},pdfauthor={Johann Schlamp, Ralph Holz, Quentin Jacquemart, Georg Carle, Ernst W. Biersack},bookmarks=true,bookmarksopen=false,pdfpagemode={UseOutlines},pdfstartview={FitV},pdfpagelayout={SinglePage},pdfborder={0 0 0},filecolor=black,linkcolor=black,urlcolor=black]{hyperref}

\newcounter{example}[section]{
\renewcommand*{\theexample}{\thesection.\arabic{example}}
\newenvironment{example}[1]{%
\centering\noindent\refstepcounter{example}\def\foo{\small Example~\theexample:~#1}}{
\vspace{-4pt}\begin{center}\foo\end{center}}

\let\orgautoref\autoref
\renewcommand{\autoref}
{\def\sectionautorefname{Section}%
\def\subsectionautorefname{Subsection}%
\def\subsubsectionautorefname{Subsection}%
\orgautoref}

\begin{document}

\title{HEAP: Reliable Assessment of\\BGP Hijacking Attacks}

\author{Johann~Schlamp,
        Ralph~Holz,
        Quentin~Jacquemart,
        Georg~Carle,
        and~Ernst~W.~Biersack%
\thanks{Manuscript received September 2, 2015; revised January 19, 2016.}%
\thanks{J.~Schlamp and G.~Carle are with the Technical University of Munich, Germany (e-mail: \{lastname\}@net.in.tum.de).}%
\thanks{R.~Holz was with Data61/CSIRO, Eveleigh, Australia. He is now with the University of Sydney, Australia (e-mail: ralph.holz@sydney.edu.au).}%
\thanks{Q.~Jacquemart and E.~W.~Biersack were with Eur\'ecom, Sophia Antipolis, France. Q.~Jacquemart is now with the University of Nice Sophia Antipolis, France (e-mail: quentin.jacquemart@unice.fr). E.~W.~Biersack is now with Caipy, Valbonne, France (e-mail: erbi@e-biersack.eu).}}

\maketitle

\begin{abstract}
The detection of BGP prefix hijacking attacks has been the focus of research
for more than a decade. However, state-of-the-art techniques fall short of
detecting more elaborate types of attack. To study such attacks, we devise a
novel formalization of Internet routing, and apply this model to routing
anomalies in order to establish a comprehensive attacker model. We use this
model to precisely classify attacks and to evaluate their impact and
detectability. We analyze the eligibility of attack tactics that suit an
attacker's goals and demonstrate that related work mostly focuses on less
impactful kinds of attacks.

We further propose, implement and test the \textit{Hijacking Event Analysis
Program (HEAP)}, a new approach to investigate hijacking alarms. Our approach
is designed to seamlessly integrate with previous work in order to reduce the
high rates of false alarms inherent to these techniques. We leverage several
unique data sources that can reliably disprove malicious intent. First, we make
use of an Internet Routing Registry to derive business or organisational
relationships between the parties involved in an event. Second, we use a
topology-based reasoning algorithm to rule out events caused by legitimate
operational practice. Finally, we use Internet-wide network scans to identify
SSL/TLS-enabled hosts, which helps to identify non-malicious events by
comparing public keys prior to and during an event. In our evaluation, we prove
the effectiveness of our approach, and show that day-to-day routing anomalies
are harmless for the most part. More importantly, we use HEAP to assess the
validity of publicly reported alarms. We invite researchers to interface
with HEAP in order to cross-check and narrow down their hijacking alerts.
\end{abstract}

\begin{IEEEkeywords}
BGP hijacking, IRR analysis, SSL/TLS measurements, routing model
\end{IEEEkeywords}

\section{Introduction} \label{sec:introduction}

The Border Gateway Protocol (BGP) is today's standard for exchanging network
routes between autonomous systems (ASes). Despite being vital to forward
traffic on the Internet, BGP does not feature security mechanisms to validate
route updates. Reports such as~\cite{study,schlamp_ccr} have shown that attacks
on BGP do occur and pose a real threat. Systems like S-BGP~\cite{Kent2000} and
RPKI~\cite{rpki} have been developed to add integrity protection and origin
authentication to BGP. However, due to the considerable resources needed to
deploy them, they are not widely used. Consequently, a number of mechanisms to
detect routing attacks have been
proposed~\cite{phas,bogus,hopcount,ispy,fingerprint,argus}.

Our contribution in this paper is twofold. First, we devise a novel
formalization of Internet routing based on concepts from formal languages. With
this model, attacks on BGP can be precisely formulated and classified according
to their effect on the global routing table. This leads us to a comprehensive
attacker model. We further discuss the motivation behind routing attacks and
learn that common prefix hijacking offers no real benefit for an attacker apart
from being destructive to the victim. More elaborate types of attacks aim to
support sustained malicious activity with benefits for an attacker such as a
chance to abuse networks to stage other attacks, send unsolicited email, or to
impersonate a victim. We show that state-of-the-art detection techniques are
not fully capable of dealing with the full spectrum of attacks. Surprisingly,
most related work focuses on the less impactful attacks, and either fully
neglects more effective and sustainable variants, or is of limited use due to a
high rate of false positives.

In the second part of this paper, we present a scheme to assess the validity of
generic hijacking alarms. This \textit{Hijacking Event Analysis Program (HEAP)}
is introduced as an automated system to reason about elaborate routing attacks.
We leverage a carefully selected set of filters proposed in previous
work~\cite{schlamp_tls,jq_moas}. With HEAP, we strive to reliably identify
\textit{legitimate} events, i.e.~hijacking alarms that are in fact false
positives.  To this end, we provide 1) administrative assurance obtained from
Internet Routing Registries, 2) operational assurance based on insights into
common routing practices, and 3) cryptographic assurance gained by
comprehensive SSL/TLS measurements.

We evaluate HEAP for a set of common routing anomalies, so-called subMOAS
conflicts, observed in BGP over the period of one month. Although a coarse
approximation of real alarms, we learn that our system is highly effective in
identifying legitimate events. We complement our findings with an instructive
case study on routing anomalies for popular networks, which host the top one
million web sites. Attackers have much interest to launch malicious activities
from such networks as they can profit from their good reputation. We show that
HEAP naturally benefits from this circumstance and yields very good results:
We are able to rule out attacks for more than 80\% of corresponding routing
anomalies.

By studying day-to-day anomalies without intrinsic evidence for an attack, we
establish a base line for our legitimization capabilities.
To demonstrate the effectiveness of HEAP, we further apply our
assessment scheme to publicly reported hijacking alarms and learn that even
such a set of highly suspicious events still contains nearly 10\% of false
positives. We find our results highly encouraging and conclude that HEAP is
suitable to assess a variety of hijacking alarms, including those of related
work that inherently exhibit high rates of false positives. Hence, we call on
researchers to continuously feed our system with their conjectural alarms.

The remainder of this paper is organised as follows. \autoref{sec:model}
presents our formalized attacker model. We discuss and assess related work
in~\autoref{sec:relatedwork}. Our methodology for HEAP is presented in
~\autoref{sec:heap}, followed by an evaluation of the approach
in~\autoref{sec:evaluation}.
\section{A Comprehensive Attacker Model} \label{sec:model}

We present a novel formalization of Internet routing based on concepts of
formal languages. With this model, routing anomalies can be precisely
expressed, classified and further assessed with respect to impact and
detectability. Previous work mostly relies on informal and often inconsistent
definitions. We believe that our routing model can improve on this situation.
It may serve as a well-founded basis for a broad spectrum of future analyses on
Internet routing. 

\subsection{Formalization of Internet Routing}

As per definition, a formal language is a set of strings of symbols constrained
by specific rules. Analagous to that, Internet routing can be represented as
the set $\mathcal{L}$ of all active BGP routes in the global routing system,
i.e.~the set of AS paths from all vantage points towards all advertised IP
prefixes. In this model, a routing attack is then defined by an attacker
extending $\mathcal{L}$ by forged routes. 

\subsubsection{Preliminaries}

Let $\Sigma_{AS}$ be the set of all ASes, $\Pi$ the set of all IP addresses, $p
\subset \Pi$ an IP prefix, and $p' \subset p$ a more specific prefix of $p$.
Let further be $(w,p) \in \Sigma_{AS}^* \times \Pi$ a route with an AS path $w
\in \Sigma_{AS}^*$, i.e.~an arbitrary concatenation of ASes, to a prefix $p \in
\Pi$, in the following denoted $r = wp$. Then, we define $\mathcal{L} \subset
\Sigma_{AS}^* \times \Pi$ as the set of \textit{active} routes to all
advertised prefixes in the global routing system, i.e.~the set of all
observable routes. $\mathcal{L}(p) \subset \mathcal{L}$ denotes the subset of
routes to a given prefix $p \subset \Pi$, such that
$$\mathcal{L}(p) \ = \ \{wuop \in \mathcal{L} \ | \ w \in \Sigma_{AS}^* \ ; \ u \in \Sigma_{AS} \ ; \ o \in \Sigma_{AS}\}$$
with $w$ being an AS subpath and $u$ the upstream AS of the origin AS $o$. For
a given route $r$ and a subprefix $p' \subset p$, we postulate $r \in
\mathcal{L}(p) \Rightarrow r \in \mathcal{L}(p')$ as a corollary, since routes
to less specific prefixes also cover more specific prefixes. Note that the
converse is false. Further, $\mathcal{L}_{P} \subset \mathcal{L}$ denotes the
set of all observable routes from a set of observation points $P \subset
\Sigma_{AS}$.
$\Pi_o$ denotes the set of IP addresses advertised by an AS $o
\in \Sigma_{AS}$, i.e.~the union of its advertised prefixes. Then, the set of
origin ASes for a prefix $p \subset \Pi$ is given by
$$O(p) \ = \ \{ o \in \Sigma_{AS} \ | \ p \subset \Pi_o \} \ .$$
Consistently, the set of origin ASes $O(p')$ for a subprefix $p' \subset p$
comprises the origin ASes for less specific prefixes such that
$$O(p') \ = \ \{ o \in \Sigma_{AS} \ | \ p' \subset \Pi_{o} \} \ = \ \bigcup_{p \ \supseteq \ p'} O(p) \ ,$$
since these ASes effectively originate routes to the particular network
$p'$.  Note again that the converse is false. The set of upstream AS neighbors
for an AS $o \in \Sigma_{AS}$ is given by
$$U(o) = \{ u \in \Sigma_{AS} \ | \ wuop \in \mathcal{L} \text{\ such that\ } w \in \Sigma_{AS}^* ; \ p \subset \Pi_o \} .$$
Due to \textit{best path selection} in BGP, the number of routes from
an observation point $s \in \Sigma_{AS}$ to a particular prefix $p \subset \Pi$
is generally bound by the number of neighboring ASes of $s$,
i.e.~$|\mathcal{L}_{s}(p)| \le |U(s)|$ holds.  In the followiong, we reuse the
unary operator $|\ .\ |$ to indicate the number of routes in a set $\mathcal{O}
\subseteq \mathcal{L}$, the length of a route $r \in \mathcal{L}$ or a subpath
$w \in \Sigma_{AS}^*$, and the number of ASes in a set $S \subseteq
\Sigma_{AS}$.

\subsubsection{Definition of Routing Attacks}

We define routing attacks as an attacker extending the global set of BGP routes
$\mathcal{L}$ by forged routes $\mathcal{F}$. Their purpose is to manipulate
existing routes in order to re-route traffic flows or to take over a victim's
Internet resources. In general, such incidents are called \textit{hijacking
attacks}. Typically, these attacks lead to topological changes in the Internet,
which can be observed by neutral BGP speakers.

Our attacker is assumed to be capable of injecting arbitrary BGP messages into the
global routing system, i.e.~he operates a BGP router and maintains a BGP
session to at least one upstream provider. We assume that the attacker is not
hindered by local filters or other validation mechanisms employed by his
upstream provider. Instead, the upstream AS indifferently redistributes all
update messages to its peers, which thus may propagate throughout the Internet.
An observation point shall be in place to monitor the propagation of BGP
messages. It is worth mentioning that data packets do not necessarily traverse
all ASes in a given path, since an attacker may craft BGP messages with a
forged AS path. Further, route updates with less attractive paths may not reach
a particular observation point due to best path selection in BGP. Without loss
of generality, an omnipresent observation pointto observe the set of all active
routes $\mathcal{L}$ is assumed for the following definitions.

In the following, we denote an attacker's AS $a \in \Sigma_{AS}$ and his
victim's AS $v \in \Sigma_{AS}$. Further, a victim's prefix is given by $p_{v}
\subset \Pi_{v}$. Then, a generic routing attack on $p_{v}$ is defined by an
attacker injecting forged routes $\mathcal{F}_{a}$ into the routing system,
such that the altered set of globally visible routes
$\hat{\mathcal{L}}(p'_{v})$ is given by
\[\hat{\mathcal{L}}(p'_{v}) \ = \ \mathcal{L}(p_{v}) \ \cup \ \mathcal{F}_{a}(p'_{v}) \text{\ \ with \ } p'_{v} \subseteq p_{v} \ .\]

\subsubsection{Impact Analysis} \label{subsec:impact}

In BGP, the impact of a hijacking attack generally depends on a best path
selection process. In particular, shorter AS paths are preferred over longer
ones, although policy-induced exceptions on a case-by-case basis may exist.
With respect to packet forwarding, routes to longer IP prefixes prevail.
Assuming the ambition to forge globally accepted routes, an attacker thus
succeeds if his routes towards a victim's network are considered best by a vast
majority of Internet participants. In practice, an attacker needs to ensure
that his bogus routes $\mathcal{F}_{a}(p'_v)$ are either
\\[-9pt]
\begin{itemize}
\item[1)] unrivaled in terms of competitive routes, i.e.~$|\mathcal{L}(p'_{v})| = 0$,\\[-9pt]
\item[2)] shortest from a global perspective, i.e.\newline$\forall r \in \mathcal{L}(p'_v), \ r_{a} \in \mathcal{F}_{a}(p'_v) \ \colon \ |r_{a}| < |r|$, or\\[-9pt]
\item[3)] more specific than all others, i.e.\newline$\forall p''_v \subseteq p'_v \subset p_v \ \colon \ \mathcal{L}(p''_v) = \mathcal{L}(p'_v) = \mathcal{L}(p_v)$.\\[-9pt]
\end{itemize}
As a consequence, the prospects of identifying an attack naturally depend on
the significance of topological changes in $\hat{\mathcal{L}}$, i.e.~on
abnormal changes to the sets of origin ASes $\hat{O}$ and upstream ASes
$\hat{U}$ for a victim's network.

\subsection{Classification of Attacks}

BGP-based attacks aim to inject falsified protocol messages into the global
routing system, which may lead to topological changes in the Internet.
Depending on the characteristics of such changes, hijacking attacks can be
classified into several subtypes with differing tactical value. 

\subsubsection{Examples}

\begin{figure}[t!]
 \centering
 \includegraphics[width=0.4825\textwidth]{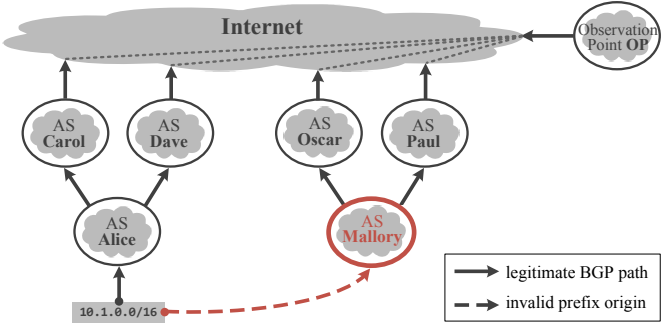}
 \vspace{-12pt}
 \caption{Origin relocation attacks in BGP.}
 \label{img:bgp_attack_surface}
 \vspace{-6pt}
\end{figure}

In addition to the formalized model, the topology in
\autoref{img:bgp_attack_surface} is used to exemplify different types of
attacks. \textit{Mallory} thereby denotes an attacker, and her autonomous
system respectively. \textit{Oscar} and \textit{Paul} are \textit{Mallory's}
upstream providers who do not validate BGP messages. The observation point
\textit{OP} receives route updates that propagate through the Internet, which
are herein after simplified to the topology-relevant attributes of BGP
messages, namely the IP prefix, also called \texttt{NLRI}, and the AS path,
refered to as \texttt{AS\_PATH}. The following expression illustrates such an
update observed at \textit{OP:}

\begin{center}
\begin{minipage}{0.4\textwidth}
\centering
\textit{OP:} \hfill $AS^{K}_{*}$ \hfill $\leftarrow$ \hfill \textit{Carol} \hfill $\leftarrow$ \hfill \textit{Alice} \hfill $\ll$ \hfill \textit{P} \hfill\quad \textit{legitimate}
\end{minipage}
\end{center}

Originating ($\ll$) at \textit{Alice}, a route update for the prefix \textit{P}
traverses \textit{Carol} and a series of $K$ ASes $AS^{K}_{*}$ to reach the
observation point \textit{OP}. For the sake of clarity, temporary convergence
effects within BGP are ignored. \textit{Alice} serves as a victim for different
kinds of attacks, \textit{Carol} and \textit{Dave} provide upstream connectivity
for \textit{Alice}.

\subsubsection{Hijacking Attacks in Practice}

Given a standard router with an already established point-to-point IP
connection to the router of an upstream provider, it is surprisingly
easy to participate in BGP and to originate an IP prefix. Consider the
following BGP configuration of \textit{Alice's} router named \texttt{rtr},
which actually represents a minimal working example:

\vspace{2pt}
\begin{center}
\begin{minipage}{0.4825\textwidth}
\footnotesize
\begin{verbatim}
rtr(config)# router bgp alice
rtr(config-router)# neighbor X.X.X.X remote-as carol
rtr(config-router)# network 10.1.0.0/16
\end{verbatim}
\end{minipage}
\end{center}
\vspace{2pt}

In this example, \textit{Alice} opens a BGP session to her neighbor router
\texttt{X.X.X.X}, which is operated by \textit{Carol}, and advertises direct
reachability of \texttt{10.1.0.0/16}. This information will be redistributed
by \textit{Carol} to her neighbors, and may eventually propagate to all BGP
routers connected to the Internet. It is as easy for an attacker to originate
arbitrary IP prefixes, irrespectively of whether legitimate routes to these IP
prefixes already exist. Depending on the specifics of the attacker's approach,
the original routes will be partially or entirely overridden in the global
routing system. Hence, this type of attack results in an \textit{origin
relocation} of a victim's network.

\subsubsection{Prefix Hijacking}\label{model:prefix_hijacking}

The most basic form of origin relocation attacks is \textit{prefix hijacking}.
An attacker thereby originates a victim's prefix at his own AS, in principle in
the same way as illustrated above. The resulting forged routes compete with the
victim's concurrent announcements. The following definition formulates this
attack scenario:
\begin{align}
\hat{\mathcal{L}}(p_{v}) =& \ \{wuvp_{v} \ | \ w \in \Sigma_{AS}^* ; \ u \in U(v)\} \,\, \cup & \text{\it\small legitimate} \label{def:prefix_hijacking} \notag \\
& \ \{wuap_{v} \ | \ w \in \Sigma_{AS}^* ; \ u \in U(a)\} & \text{\bf\small forged} \notag \\
\text{\small with\quad} \hat{O}(p_{v}) =& \ O(p_{v}) \ \cup \ \{a\} \notag \\
\text{\small and\quad} \hat{U}(v) =& \ U(v) \notag
\end{align}
The set of origin ASes $\hat{O}(p_{v})$ for the prefix $p_{v}$ now comprises
two ASes, while the set of the victim's upstream ASes $\hat{U}(v)$ remains
unchanged. In literature, a situation with $|O(p)| >\nobreak 1$ is often called
a \textit{multi-origin AS (MOAS)}. Given the exemplary topology in
\autoref{img:bgp_attack_surface}, \textit{Mallory} may craft a BGP update
message as listed below. At the same time, legitimate paths advertised by
\textit{Alice} are present in the global routing table.

\vspace{-7.5pt}
\begin{example}{Prefix Hijacking.}
\begin{center}
\begin{minipage}{0.48\textwidth}
\centering
\small
$OP:$ \hfill $AS^{K}_{*}$ \hfill $\leftarrow$ \hfill \textit{Carol} \hfill $\leftarrow$ \hfill \textit{Alice} \hfill $\ll$ \hfill \texttt{10.1.0.0/16} \hfill\quad \textit{legitimate} \\[3.5pt]
$OP:$ \hfill $AS^{L}_{*}$ \hfill $\leftarrow$ \hfill \textit{Dave} \hfill $\leftarrow$ \hfill \textit{Alice} \hfill $\ll$ \hfill \texttt{10.1.0.0/16} \hfill \textit{legitimate} \hfill \\[2pt]
$OP:$ \hfill $AS^{M}_{*}$ \hfill $\leftarrow$ \hfill \textit{Oscar} \hfill $\leftarrow$ \hfill \fbox{\textit{Mallory}} \hfill $\ll$ \hfill \texttt{10.1.0.0/16} \hfill \textbf{forged} \hfill \\[2pt]
\end{minipage}
\end{center}
\end{example}

As shorter AS paths are generally preferred over longer ones, the attack is
likely to succeed for observation points $s$ where $M < \text{min}(K,L)$ holds,
and for
\[\{ s \in \Sigma_{AS} \ | \ \forall w_{v}p_{v} \in \mathcal{L}_s(p_{v}) \ \exists \ w_{a}p_{v} \in \mathcal{F}_a(p_{v}) \colon |w_{a}| < |w_{v}| \}\]
respectively. However, it is safe to assume that clients that are topologically
close to \textit{Carol} or \textit{Dave} still reach the victim \textit{Alice},
since shorter legitimate routes take precedence. The Internet thus decomposes
into two disjoint parts: one part that is affected by the forged announcement,
in literature often called the \textit{poisoned} part, and one that remains
unaffected. 

\subsubsection{Subprefix Hijacking} \label{subsec:subprefix}

To overcome the limited impact inherent to prefix hijacking, an attacker can
leverage longest prefix matching in IP routing with so-called \textit{subprefix
hijacking}. To this end, the attacker originates a subprefix $p'_{v} \subset
p_{v}$ at his AS, thereby injecting a new set of routes
$\mathcal{F}_{a}(p'_{v})$ into the global routing system as given by:
\begin{align}
\hat{\mathcal{L}}(p_{v}) =& \ \{wuvp_{v} \ | \ w \in \Sigma_{AS}^* ; \ u \in U(v)\} \,\, \cup & \text{\it\small legitimate} \label{def:subprefix_hijacking} \notag \\
& \ \{wuap'_{v} \ | \ w \in \Sigma_{AS}^* ; \ u \in U(a)\} & \text{\bf\small forged} \notag \\
\text{\small with\quad} \hat{O}(p_{v}) =& \ O(p_{v}), \ \hat{O}(p'_{v}) = O(p_{v}) \, \cup \, \{a\} \notag \\
\text{\small and\quad} \hat{U}(v) =& \ U(v) \notag
\end{align}
\noindent Subprefix hijacking attacks generally have global impact, since
routes to the more specific prefix $p' \subset p$ dominate.
Such incidents with $|O(p)| > 0$ and $|O(p') \setminus
O(p)| > 0$ are called \textit{subprefix multi-origin AS (subMOAS)}.

Note that the victim might readily advertise routes to a prefix and a
corresponding subprefix concurrently to the attacker's subprefix route,
i.e.~condition 3) in \autoref{subsec:impact} does not hold since
$\mathcal{L}(p_v) \subset \mathcal{L}(p'_v)$. In this case, the event can also
be considered a MOAS event. Otherwise, it is called a \textit{strict} subMOAS,
and subprefix hijacking respectively. We assume this variant in the following.
While virtually all Internet participants are affected by strict subprefix
hijacking, it may be tempting to conclude that only part of the victim's
network, i.e.~subprefixes, can be taken over. As a matter of fact, this is not
the case. An attacker can easily craft multiple update messages such that the
set of forged routes $\mathcal{F}_{a}(p_{v})$ fully covers the prefix $p_{v}$
with more specific routes:
$$\mathcal{F}_{a}(p_{v}) \ = \ \bigcup_{p'_v \in \Pi} \mathcal{F}_{a}(p'_v) \ \text{such that\ \ } p_{v} \ = \ \bigcup p'_v \ .$$
With respect to \autoref{img:bgp_attack_surface}, \textit{Mallory} could thus
inject the following BGP routes:

\vspace{-7.5pt}
\begin{example}{Subrefix Hijacking.}
\begin{center}
\begin{minipage}{0.48\textwidth}
\centering
\small
$OP:$ \hfill $AS^{K}_{*}$ \hfill $\leftarrow$ \hfill \textit{Carol} \hfill $\leftarrow$ \hfill \textit{Alice} \hfill $\ll$ \hfill \texttt{10.1.0.0/16} \hfill \textit{legitimate} \hfill \\[7pt]
$OP:$ \hfill $AS^{L}_{*}$ \hfill $\leftarrow$ \hfill \textit{Dave} \hfill $\leftarrow$ \hfill \textit{Alice} \hfill $\ll$ \hfill \texttt{10.1.0.0/16} \hfill \textit{legitimate} \hfill \\[5pt]
$OP:$ \hfill $AS^{M}_{*}$ \hfill $\leftarrow$ \hfill \textit{Oscar} \hfill $\leftarrow$ \hfill \fbox{\textit{Mallory}} \hfill $\ll$ \hspace{0pt} \hfill \texttt{10.1.0.0/17} \hfill \textbf{forged} \hfill \\[1.5pt]
$OP:$ \hfill $AS^{M}_{*}$ \hfill $\leftarrow$ \hfill \textit{Oscar} \hfill $\leftarrow$ \hfill \fbox{\textit{Mallory}} \hfill $\ll$ \hfill \texttt{10.1.128.0/17} \hfill \textbf{forged} \hfill \\[2pt]
\end{minipage}
\end{center}
\end{example}

By advertising a victim's network with BGP updates split up into multiple
longer prefixes, as given by \texttt{10.1.0.0/17} and \texttt{10.1.128.0/17} in
the example above, subprefix hijacking can be as extensive as regular prefix
hijacking, with the advantage of globally preferred routes at the same time.
Notwithstanding this possibility, an attacker might be satisfied with hijacking
individual subnets of high value only.

\subsubsection{Other Types of Attack}

Origin relocation attacks as discussed above lead to noticeable changes in the
set of origins for a victim's prefix. In contrast, \textit{route diversion}
attacks aim at the manipulation of AS paths towards a victim. In its basic
form, an attacker hijacks a victim's prefixes and its AS, which effectively
disguises the attack by hiding the attacker's own AS. Moreover, tailored
attacks can be derived to impersonate a victim from an administrative point of
view by stealthily hijacking abandoned Internet resources. We have studied such
\textit{hidden takeover} attacks in great detail in previous
work~\cite{schlamp_ccr,schlamp_resources}. An even more sophisticated attack
based on AS path manipulation aims at stealthily intercepting a victim's
traffic while maintaining the victim's connectivity. Such
\textit{man-in-the-middle} attacks require a stable backhaul link from the
attacker to the victim to forward eavesdropped packets. Interestingly,
BGP itself, and its loop prevention mechanism respectively, can be leveraged to
accomplish this task. Our routing model allows to formalize and study
these types of attacks. For this paper, however, we consider them beyond
scope.

\subsection{Motivation behind Hijacking Attacks} \label{subsec:motivation}

Hijacking attacks may serve a variety of purposes, which can be divided into
\textit{destructive} and \textit{abusive} variants. An obvious intent is to
inflict damage to a victim's operations by disrupting network connectivity,
which is often called \textit{blackholing} in literature. Furthermore, hijacked
networks can be abused for malicious short-term activities, for instance to
launch a fast, temporary, and massive spam campaign. Abusive actions
may aim at hosting illegal services, like phishing web sites, or serve to
establish a stable base of operations for subsequent attacks, e.g. to command
botnets from a safe-house network. More sophisticated attacks aim at
compromising a victim's reputation by carrying out illicit actions.
Lastly, attacks can be tailored to break confidentiality or integrity of a
particular victim's communications by means of interception.

Prefix hijacking attacks \textit{partially} disrupt a victim's connectivity,
but are of limited use in other respects, like hosting malicious services,
since a significant part of the Internet might still prefer the victim's
routes. In contrast, subprefix hijacking attacks are capable of breaking
communications entirely, i.e. all hosts inside the victim's network become
globally unreachable. Hence, this type of attack is also useful for launching
illegal operations from a hijacked network, and an important element for more
sophisticated attacks like AS hijacking or man-in-the-middle interception.
Having no real-world benefit apart from partially disconnecting a victim's
hosts from the Internet, it is surprising to see that state-of-the-art focuses
primarily on prefix hijacking.
\section{Related Work} \label{sec:relatedwork}

There is a large body of literature on the detection of BGP-based routing
attacks. Corresponding techniques can be divided into control-plane and
data-plane techniques, with hybrid approaches emerging recently.

\subsection{State-of-the-art Detection}

Possibly the first attempt to detect hijacking attacks was presented with
PHAS~\cite{phas}. It is a control-plane technique focusing exclusively on
reporting MOAS conflicts. In an effort to reduce high rates of false positives,
PHAS utilizes an adaptive time-window that prevents recurring alerts. The
authors of \cite{bogus} provide further heuristics to assess MOAS conflicts with
respect to compliance with economy-based routing policies. Careful tuning of
heuristic parameters is necessary to yield suitable detection results. The
approach thus tends to reduce false alarms at the cost of an increased rate of
false negatives. LOCK~\cite{lock} offers to pinpoint attackers in the AS-level
topology. The Buddyguard system~\cite{buddy} uses a learning-based approach to
detect abnormal routing changes. The creators of BGPmon.net~\cite{bgpstream}
provide MOAS and subMOAS alarms in real-time, but do not publish details about
their methodology.

In~\cite{hopcount}, a light-weight distributed measurement scheme (LWDS) is
proposed. This data-plane technique is based on the assumption that path
measurements yield stable path lengths for a majority of networks. Major
violations of this conjecture hint at a suspicious change in network location
if a set of reference points close to this network remains unaffected. Due to
its dependency on suitable vantage points and reference nodes, it is difficult
to deploy the scheme on a larger scale. A comparable technique that leverages
latency measurements from multiple vantage points is proposed in\,\cite{ping}.
The StrobeLight system\,\cite{strobelight} detects hijacking
attacks in a similar way, though on a per-operator basis only. A more popular
operator-centric approach is given with iSpy\,\cite{ispy}. By carrying out
data-plane measurements from an operator's network to major transit ASes,
attacks are detected based on specific outage patterns that reflect the
partitioning of the Internet into affected and unaffected parts (refer to
\autoref{model:prefix_hijacking}). iSpy cannot differentiate between subprefix
hijacking and temporary link failures or congestion near the victim's network.

\begin{table}[t!]
 \centering
 \footnotesize
 \renewcommand{\arraystretch}{1.12}
 \setlength{\tabcolsep}{1.75pt}
 \begin{tabularx}{0.485\textwidth}{|X|cc|ll|c|}
  \cline{2-6}
  \multicolumn{1}{c|}{} & \bf \footnotesize Prefix & \bf \footnotesize Subprefix & \multicolumn{2}{c|}{\multirow{2}{*}{Applicability}} & Public\\[-2pt]
  \multicolumn{1}{c|}{} & \bf \footnotesize Hijacking & \bf \footnotesize Hijacking & & \multicolumn{1}{c|}{} & Interface\\
  \hline
  \multicolumn{6}{|l|}{\it control-plane detection} \\
  \hline
  \ \ PHAS\,\cite{phas} & + & o & \ \ o & \it\footnotesize (operator-level) & no \\
  \ \ Bogus Routes\,\cite{bogus} & + & o & \ \ + & \it\footnotesize (global scale) & no \\
  \ \ BGPmon.net\,\cite{bgpstream} & x & x & \ \ + & \it\footnotesize (global scale) & yes \\
  \hline
  \multicolumn{6}{|l|}{\it data-plane detection} \\
  \hline
  \ \ LWDS\,\cite{hopcount} & + & + & \ \ -- & \it\footnotesize (incident-level) & no \\
  \ \ iSpy\,\cite{ispy} & + & -- & \ \ o & \it\footnotesize (operator-level) & no \\
  \hline
  \multicolumn{6}{|l|}{\it hybrid techniques} \\
  \hline
  \ \ Fingerprints\,\cite{fingerprint} & + & o & \ \ + & \it\footnotesize (global scale) & no \\
  \ \ Argus\,\cite{argus} & + & o & \ \ + & \it\footnotesize (global scale) & yes \\
  \hline\hline
  \ \ \bf HEAP (our system) & \bf o & \bf ++ & \ \ \bf + & \it\footnotesize (global scale) & yes \\[0.5pt]
  \hline
  \multicolumn{6}{c}{} \\[-8pt]
  \multicolumn{6}{c}{\footnotesize ++/+ \, practical usefulness \quad o \, theoretic applicability} \\
  \multicolumn{6}{c}{\footnotesize -- \, not supported \quad x \, no details published} \\
 \end{tabularx}
 \caption{Comparison of popular state-of-the-art detection systems.}
 \vspace{-3pt}
 \label{tab:detection_systems}
\end{table}

One of the first hijacking detection systems to include both passive and active
measurements is a fingerprint-based approach\,\cite{fingerprint}. This system
utilizes a monitoring scheme for BGP combined with active measurements
to identify two disjoint parts of the Internet as outlined above. For subprefix
hijacking, where no such partitioning occurs (see \autoref{subsec:subprefix}),
several heuristics are proposed. A similar approach is realized with the Argus
system\,\cite{argus}. This framework detects anomalies in BGP and confirms
hijacking attacks by carrying out measurements from public looking glasses to
identify a poisoned part of the Internet. The authors of\,\cite{idle} extend
the set of active scans, while~\cite{characteristics} proposes additional types
of fingerprints. A complementary approach to assess immediate effects of
control-plane anomalies onto the data-plane is to perform measurements during
and after an event\,\cite{spamtracer}.

\subsection{Assessment and Comparison}

Looking at \autoref{tab:detection_systems}, it is surprising to see that
related work concentrates its effort on the detection of prefix hijacking
attacks, despite the fact that this kind of attack is of very limited use for
an attacker in practice. As a matter of fact, subprefix hijacking attacks offer
more useful means for a broader area of operations, but are mostly neglected in
state-of-the-art techniques. HEAP, our own contribution, aims to improve on
this situation by providing reliable means to assess alarms relating to this
specific type of attack.
\section{HEAP: A Real-time Framework} \label{sec:heap}

We propose a novel \textit{Hijacking Event Analysis Program (HEAP)} to
reliably assess hijacking alarms. We combine techniques
from prior work~\cite{schlamp_tls,jq_moas} to identify incidents with
legitimate cause. Our goal is \textit{not} to develop a new alarm system.
Instead, HEAP is designed to receive input from available detection systems
to reduce their rate of false alarms.

\subsection{System Architecture} \label{subsec:arch}

HEAP leverages three distinct data sources to assess hijacking events. Our main
assumption here is that an attacker is capable to hijack networks in BGP, but
cannot alter orthogonal data sources that relate to the operation of hijacked
networks. Hence, we are able to rule out an attack if these data sources
legitimize a suspicious routing anomaly. First, Internet Routing Registries
(IRR) are utilized to infer legitimate business relationships between an
attacker and his alleged victim. Second, a topological analysis is carried out
in order to identify benign anomalies resulting from common operational
practices. And lastly, SSL/TLS measurements yield cryptographic assurance that
traffic to a supposedly hijacked network is still routed to the alleged victim.

\begin{figure}[t!] \centering
 \includegraphics[width=0.4885\textwidth]{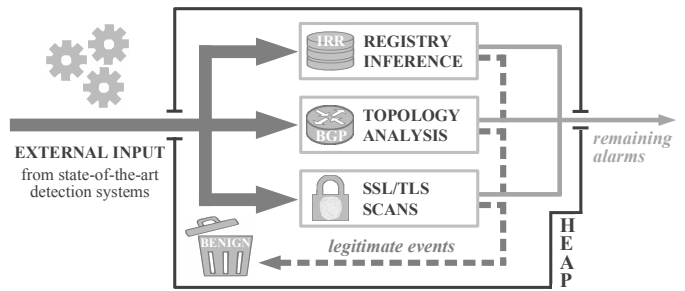}
 \vspace{-10pt}
 \caption{The Hijacking Event Analysis Program (HEAP).}
 \label{fig:heap}
 \vspace{-3pt}
\end{figure}

\autoref{fig:heap} illustrates the main workflow within HEAP. Given external
alarms fed into the system, legitimate events are identified and eliminated as
false positives based on the aforementioned data sources. Note that the SSL/TLS
component needs a tight coupling to external systems that provide us with alarms
since corresponding scans have to be carried out in response to the input
received. The remaining events are highly suspicious indications for an
attack, even if we take into account that we cannot make further assumptions
about their nature. This is for two reasons: 1) the input source already
provides potential hijacking candidates, and 2) none of our filter techniques
yields evidence for a legitimate cause. In the following, we will show that
this is indeed improbable for benign events. The remaining alarms lend
themselves well to manual inspection, with a rich set of background information
readily available from the individual analysis steps.

\subsection{Filtering Methodology}

All filters applied by HEAP are executed concurrently. HEAP is easily
extensible, i.e.~additional filters can be incorporated without difficulty.
Three independent techniques to eliminate legitimate alarms have
been implemented so far.

\subsubsection{Utilizing IRR Databases}\label{subsec:irr}

\definecolor{rirgray}{rgb}{0.75,0.75,0.75}

\begin{table*}[t!]
 \setlength{\tabcolsep}{4.0pt}
 \footnotesize
 \centering
 \renewcommand{\arraystretch}{0.75}
 \begin{tabularx}{0.9\textwidth}{Xrr|rr|rr|rr|rr}
  & \multicolumn{2}{c|}{\bf AfriNIC} & \multicolumn{2}{c|}{\bf APNIC} & \multicolumn{2}{c|}{\bf $^{1}$ARIN} & \multicolumn{2}{c|}{\bf LACNIC} & \multicolumn{2}{c}{\bf RIPE} \\
  \multicolumn{1}{l}{\bf instance} & \multicolumn{1}{c}{nodes} & \multicolumn{1}{c|}{relations} & \multicolumn{1}{c}{nodes} & \multicolumn{1}{c|}{relations} & \multicolumn{1}{c}{nodes} & \multicolumn{1}{c|}{relations} & \multicolumn{1}{c}{nodes} & \multicolumn{1}{c|}{relations} & \multicolumn{1}{c}{nodes} & \multicolumn{1}{c}{relations} \\
  \toprule
  \texttt{MNTNER} & \bf 2,624 & \textcolor{rirgray}{---} & \bf 20,129 & \textcolor{rirgray}{---} & \bf n/a & \textcolor{rirgray}{---} & \bf n/a & \textcolor{rirgray}{---} & \bf 53,670 & \textcolor{rirgray}{---} \\
  $\leftarrow$\textit{maintained\_by}\textbf{--} \texttt{[*]} & \textcolor{rirgray}{---} & 133,186 & \textcolor{rirgray}{---} & 1,919,397 & \textcolor{rirgray}{---} & n/a & \textcolor{rirgray}{---} & n/a & \textcolor{rirgray}{---} & 5,620,385 \\
  \midrule
  \texttt{ORGANISATION} & \bf 1,877 & \textcolor{rirgray}{---} & \bf n/a & \textcolor{rirgray}{---} & \bf 2,976,707 & \textcolor{rirgray}{---} & \bf n/a & \textcolor{rirgray}{---} & \bf 90,102 & \textcolor{rirgray}{---} \\
  $\leftarrow$\textit{org}\textbf{--} \texttt{[*]} & \textcolor{rirgray}{---} & 32,476 & \textcolor{rirgray}{---} & n/a & \textcolor{rirgray}{---} & 3,536,502 & \textcolor{rirgray}{---} & n/a & \textcolor{rirgray}{---} & 249,319 \\
  \midrule
  \texttt{AUT-NUM} & \bf 1,239 & \textcolor{rirgray}{---} & \bf 9,485 & \textcolor{rirgray}{---} & \bf 24,939 & \textcolor{rirgray}{---} & \bf 5,193 & \textcolor{rirgray}{---} & \bf 29,206 & \textcolor{rirgray}{---} \\
  $\leftarrow$\textit{origin}\textbf{--} \texttt{ROUTE} & \textcolor{rirgray}{---} & 464 & \textcolor{rirgray}{---} & 216,865 & \textcolor{rirgray}{---} & $^{2}$583,296 & \textcolor{rirgray}{---} & n/a & \textcolor{rirgray}{---} & 279,532 \\
  $\leftarrow$\textit{import}\textbf{--} \texttt{AUT-NUM} & \textcolor{rirgray}{---} & 6 & \textcolor{rirgray}{---} & 10,734 & \textcolor{rirgray}{---} & n/a & \textcolor{rirgray}{---} & n/a & \textcolor{rirgray}{---} & 228,509 \\
  \midrule
  \texttt{INETNUM} & \bf 85,672 & \textcolor{rirgray}{---} & \bf 924,584 & \textcolor{rirgray}{---} & \bf 2,910,623 & \textcolor{rirgray}{---} & \bf 342,104 & \textcolor{rirgray}{---} & \bf 3,995,522 & \textcolor{rirgray}{---} \\
  \midrule
  \texttt{ROUTE} & \bf 443 & \textcolor{rirgray}{---} & \bf 97,858 & \textcolor{rirgray}{---} & \bf $^{2}$600,940 & \textcolor{rirgray}{---} & \bf n/a & \textcolor{rirgray}{---} & \bf 267,216 & \textcolor{rirgray}{---} \\
  \bottomrule
 \end{tabularx}
 \caption{Data stored in our graph database. August, 2015.}
 \centering
 {\footnotesize $^{1}$~ARIN's object identifiers can be directly mapped to RIPE's schema (e.g.~\texttt{ASHandle} $\rightarrow$ \texttt{AUT-NUM}).\\[1pt]}
 {\footnotesize $^{2}$~Implicitly given in ARIN's \texttt{INETNUM} objects (via \texttt{OriginAS} attributes).}
 \label{table:ripedb}
\end{table*}

Regional Internet Registrars (RIRs) maintain so-called Internet Routing
Registries (IRR), i.e.~databases that contain information pertaining to the
management of Internet resources. A recent study~\cite{ccrirr} matched prefixes
and ASes observed in BGP and IRRs by looking for appropriate database objects.
We provide a more generalized set of inference rules to identify benign routing
events, which take into account multiple prefix origins observed in BGP as well
as complex relationships between affected prefixes and suspicious ASes. The
fundamental assumption behind our approach is that an attacker does not have
the credentials to change an IRR database in order to cover his attack. To
disprove an attack, we accordingly look for legitimizing database relations
between the entities involved in a hijacking alarm, e.g. for a common
organisation referenced by two ASes. To this end, we download and evaluate
snapshots of the IRR databases, which are provided by RIRs on a daily basis. We
use a graph database to store the extracted information using the schema
presented in \autoref{table:ripedb} and track all changes over time.  Note that
IRR databases are updated by individual resource holders and can thus be
outdated or even hold conflicting information. Our filter accounts for this by
strictly searching for legitimizing relationships without drawing any
conclusions in their absence.

\begin{figure}[t!]
 \centering
 \includegraphics[width=0.45\textwidth]{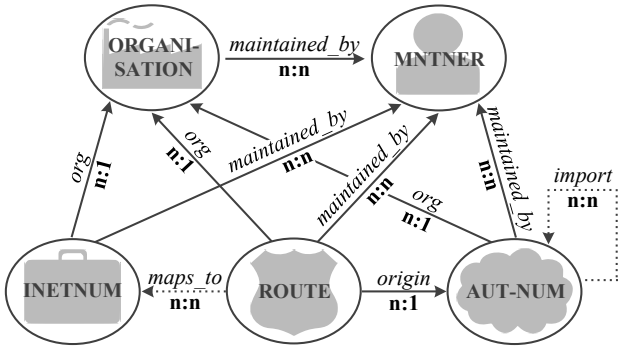}
 \vspace{2pt}
 \captionof{figure}{Relevant data objects and relations in IRR databases.}
 \label{fig:dbmodel}
 \vspace{-6pt}
\end{figure}

To assess a given hijacking alarm, we map the affected AS numbers and prefixes
to resource objects, i.e.~graph nodes, in our graph database. We then traverse
the graph along a path of legitimizing relations that document a right to use,
which are given by \texttt{AUT-NUM} and \texttt{INETNUM} objects linked by
\textit{import}, \textit{origin}, \textit{maintained\_by} or
\textit{org} relations. We look for such paths between a) two affected ASes, or
b) a prefix and its origin AS. If we succeed with a), we can infer a valid
business relationship between the victim and the suspected attacker. If we
succeed with b), the suspected attacker holds ownership rights for the prefix
and is thus authorized to originate the prefix from his AS. Compared to our
previous work~\cite{schlamp_tls}, we extended this filter to support the IRR
databases provided by all five RIRs. Note that we transform all IRR databases
into the RIPE data model as presented in \autoref{fig:dbmodel}, since it is
most consistent and represents a superset of all available information. In
fact, AfriNIC and APNIC already store their data in a similar format and can
thus be directly processed by our filter. LACNIC and ARIN utilize their own
data models, which can be converted in spite of some missing data points.

RIPE-based IRR databases model access rights with the help of \texttt{MNTNER}
objects. Only maintainers with valid credentials can modify or delete objects.
For any object, this is expressed by adding a \textit{maintained\_by} reference
pointing to the respective \texttt{MNTNER} object. \texttt{ORGANISATION}
objects are mainly used to provide administrative contact details. For privacy
reasons, most IRR database snapshots do not include details, but unique
references to these objects are preserved. \texttt{INETNUM} objects document
allocated or assigned IPv4 prefixes managed by the respective RIR.
\texttt{AUT-NUM} objects represent AS numbers and may be referenced as the
\textit{origin} of \texttt{ROUTE} objects. Such \texttt{ROUTE} objects are
created by resource holders and are used to document or confirm intended prefix
announcements by specific ASes. To create a \texttt{ROUTE} object, the resource
holder needs to provide valid credentials for the respective \texttt{INETNUM}
and \texttt{AUT-NUM} objects. A corresponding \textit{maps\_to} relation is
derived by our parsing algorithm, as is the case with \textit{import} relations
deduced from free-text description fields, which are often used to model
routing policies in the so-called Routing Policy Specification Language (RPSL).
When resources are deleted from a database, RPSL definitions may still
reference (now) non-existing ASes. We account for this by tracking such
orphaned \textit{import} relations.

\autoref{table:ripedb} provides further details on selected objects that are
relevant to our approach. Our combined database holds more than 15 million
nodes and 45 million relations extracted from the five individual IRR
databases. Entries marked with \textit{n/a} are not available in the respective
database snapshots. For the RIPE database, for instance, we can see that less
than 55,000 \texttt{MNTNER} objects share more than 5 million incoming
\textit{maintained\_by} references. Although optional, roughly
90,000 \texttt{ORGANISATION} objects are referenced by 250,000 other objects.
About 280,000 \texttt{ROUTE} objects bind prefix announcements to less than
30,000 \texttt{AUT-NUM} objects. Furthermore, these \texttt{AUT-NUM} objects
document nearly 230,000 \textit{import} routing policies. We will see that this
rich information allows our filter to be highly effective---except in the case
of LACNIC, where none of the necessary cross-referencing objects are provided
in the daily database snapshots.

\subsubsection{Topology Reasoning} \label{subsec:topo}

The next filter to legitimize routing anomalies is a topology-based reasoning
algorithm. The key idea is that an attacker is unlikely to hijack his own
upstream provider. This assumption is based on the fact that the attacker's
malicious BGP updates need to propagate via this upstream provider, who could
simply counter an attack by filtering them out. As a consequence, we can rule
out an attack if the suspected attacker resides in the downstream AS path of
his victim.

To identify such benign anomalies, we utilize BGP collectors to extract all AS
paths that lead to the affected prefixes and ASes respectively. If we do not
find any AS path that contains both the attacker's and the victim's AS, we
cannot draw any further conclusions. The same is true for AS paths in which the
attacker is located upstream of his victim. In contrast, we can infer a
legitimate cause of the anomaly if the attacker is actually located downstream
of the victim, i.e.~if we find a particular AS path in which the victim's AS
precedes the attacker's AS. In this case, we can rule out malicious intent.

Such benign situations might occur, for instance, if smaller organizations
obtain Internet connectivity and an IP prefix from a larger carrier. Other
reasons can be static routes invisible to BGP, imperfect multihoming setups, or
even misconfiguration. In our evaluation, we will see that a significant part
of day-to-day routing anomalies is caused by such topological constellations.

\subsubsection{Cryptographic Assurance with SSL/TLS} \label{subsec:tls}

We use a final strong filter that is based on our regular Internet-wide scans
of SSL/TLS protocols (refer to~\cite{holzndss2016} for further details). For
any given hijacking alarm concerning a certain IP prefix, we verify if affected
SSL/TLS hosts present the same public key before and during the event. We make
the assumption that an attacker cannot gain access to the private keys of a
victim's hosts, and thus cannot perform successful SSL/TLS handshakes. We
conclude that such cases cannot be attacks.

A prerequisite for this filter is a \textit{ground truth scan} to obtain a
known-correct mapping from IP addresses to public keys that are used on
corresponding machines. Given such a ground truth data set, we can carry out
\textit{validation scans} to hosts in a prefix that relates to a hijacking
alarm and compare the retrieved public keys. Note that it is
imperative for these scans to be executed in a timely manner, i.e.~we need to
compare public keys during the life time of an event. A tight coupling to the
alarming system is dispensable if we can retroactively ascertain that an event
lasted for the entire duration of a corresponding scan. Since our system is
designed to assess subprefix hijacking attacks, which affect the Internet as a
whole (refer to \autoref{subsec:impact}), the vantage point for our SSL/TLS
measurements can be chosen freely. For this paper, we employed a scanning
machine hosted at our university in Munich (AS56357).

Compared to our previous work~\cite{schlamp_tls}, we greatly extended the
ground truth scans to a variety of popular SSL/TLS-based protocols.
\autoref{table:tlshosts} shows all scans that were carried out for this work.
In many cases, we scanned for both TLS and STARTTLS, a common extension to
network protocols that allows for opportunistic use of TLS. It is instructive
to see that the use of TLS varies greatly between application-layer
protocols. An open, dedicated port does not imply support for TLS per se. Due
to the marginal contributions that our scans of XMPPS and IRCs provided, we
have since stopped scanning these protocols.

\newcolumntype{R}{>{\raggedleft\arraybackslash}X}
\begin{table}[t]
 \footnotesize
 \setlength{\tabcolsep}{2.75pt}
 \renewcommand{\arraystretch}{1.12}
 \centering
 \begin{tabularx}{0.4885\textwidth}{lRrrrr}
 & port & time & port open & handshake & in \% \\
 \toprule
 \textbf{Implicit SSL/TLS} & & 27d & \textbf{72,546,563} & \textbf{38,146,816} & \textbf{52.58\%} \\
 \midrule
 HTTPS & 443 & 10d & 42,676,912 & 27,252,853 & 63.85\% \\
 SMTPS & 465 & 2d & 7,234,817 & 3,437,382 & 47.51\% \\
 IMAPS & 993 & 3d & 6,297,805 & 4,121,108 & 65.43\% \\
 POP3S & 995 & 3d & 5,186,724 & 2,797,300 & 53.93\% \\
 FTPS & 990 & 2d & 2,657,680 & 344,400 & 12.95\% \\
 LDAPS & 636 & 2d & 2,273,771 & 112,978 & 4.96\% \\
 XMPPS/CLIENT & 5223 & 2d & 2,223,994 & 70,441 & 3.16\% \\
 XMPPS/SERVER & 5270 & 1d & 2,046,204 & 1,693 & 0.08\% \\
 IRCS & 6697 & 2d & 1,948,656 & 8,661 & 0.44\% \\
 \midrule
 \textbf{Explicit SSL/TLS} & & 9d & \textbf{51,768,705} & \textbf{18,316,920} & \textbf{35.38\%} \\
 \midrule
 FTP/STARTTLS & 21 & 2d & 14,493,966 & 2,939,048& 20.27\% \\
 SMTP/STARTTLS & 25 & 2d & 12,488,000 & 3,848,843 & 30.82\% \\
 POP3/STARTTLS & 110 & 1d & 8,930,688 & 4,074,211 & 45.62\% \\
 IMAP/STARTTLS & 143 & 2d & 8,006,617 & 4,076,809 & 50.91\% \\
 SUBMISSION/STARTTLS & 587 & 2d & 7,849,434 & 3,378,009 & 43.03\% \\
 \midrule
 \textbf{Total SSL/TLS scans} & & 36d & \textbf{124,315,268} & \textbf{56,463,736} & \textbf{45.42\%} \\
 \bottomrule
 \end{tabularx}
 \vspace{2pt}
 \caption{Scanned SSL/TLS hosts for our ground truth data set. All measurements were carried out in July, 2015.}
 \label{table:tlshosts}
\end{table}

On total, we tried to open connections to 124,315,268 individual ports. For
successful SSL/TLS handshakes, we downloaded the certificate and extracted the
public key. Note that we only consider keys that were unique across the whole
dataset. This precaution eliminates the risk of falsely legitimizing events
imposed by default certificates. Such certificates often ship with popular web
server software or with SSL/TLS-enabled devices, and could thus be presented by an
attacker as well. We relax this condition only where the same key is presented
by a single host for multiple protocols. This finally yields a total of
12,800,474 available keys, which were presented by a total of 8,402,023
different hosts.

\subsection{Applicability} \label{subsec:applicability}

Our approach works best for the assessment of subprefix hijacking alarms.
Attacks that build upon the manipulation of AS paths, like AS hijacking, for
instance, can be assessed with HEAP as well. Due to a general lack of initially
suspicious input events, however, we exclude this kind of attack from our
analysis. 

Ordinary prefix hijacking attacks, and MOAS conflicts respectively, impose
limitations to our SSL/TLS filter, since we cannot assure that measurements reach
a supposedly hijacked network. With the Internet decomposing into two disjoint
parts as discussed in \autoref{model:prefix_hijacking}, our SSL/TLS scans might
reach either part, which prevents reliable conclusions. Nevertheless, we can
deactivate the SSL/TLS filter for such cases. BGP-based man-in-the-middle
attacks~\cite{defcon} are especially hard to identify~\cite{blackhat}. In an
interception scenario, in which an attacker is able to forward our active scans
to the victim, the SSL/TLS filter would wrongly legitimize the incident. Hence,
these attacks are left for future studies.

We acknowledge that our approach depends on external input, thus it is arguably
not a full-fledged detection system. In our evaluation, however, we show that
we arrive at remarkable validation results even for a superset of potential
alarms.
\section{Evaluation} \label{sec:evaluation}

Most of the proposed detection techniques in previous work (see
\autoref{tab:detection_systems}) do not offer publicly available interfaces
yet. We compensate for the resulting lack of real alarms by studying common
subMOAS conflicts observed in BGP. Such cases occur numerous times per day and
do not indicate attacks per se. Although more careful heuristics should be
employed in practice to actually feed suspicious events into HEAP, we are able
to establish a base line for its validation capabilities nonetheless. We further
use HEAP to cross-check a set of publicly reported real
hijacking alarms and demonstrate its practical usefulness in identifying false
positives.

\subsection{Experiment Setup} \label{subsec:setup}

\begin{figure}[t!]
 \centering
 \vspace{2.5pt}
 \includegraphics[width=0.405\textwidth]{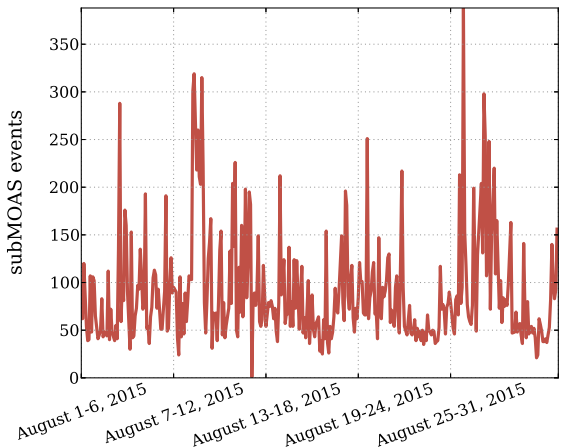}
 \caption{subMOAS events observed with our experiment.}
 \label{fig:subtimes}
 \vspace{-3pt}
\end{figure}

Our evaluation setup comprises several steps that are repeatedly executed.
First, we obtain a full BGP table export holding all prefixes currently present
in the global routing system and construct a binary prefix tree such that a
tree node holds the date and origin AS of an announcement. To discover emerging
subMOAS events, we obtain subsequent BGP messages and update the binary tree
accordingly. We consequently extract all \textit{strict} subMOAS events (refer
to \autoref{subsec:subprefix}) from the tree that newly appeared in the BGP
updates. For these, we apply our filters individually, i.e.~we query our graph
database for business and resource relations, construct an event-specific
AS-level topology, and initiate SSL/TLS measurements for affected hosts that
also appear in our ground truth data set. We then retrieve the scan results
from successful SSL/TLS handshakes and compare the cryptographic host keys with
those from our initial ground truth scan. As outlined earlier, we need to
ensure that our scans actually reached a targeted prefix, since our BGP view,
respectively the subMOAS events, might be outdated at the time of observation.
Thus, we re-evaluate the aforementioned BGP update messages and discard
previous scan results for which a subMOAS event changed or vanished during a
scan. Note that we accordingly sanitize the data in our ground truth, too, to
ensure that no initially scanned SSL/TLS hosts were affected by subMOAS events.
This led to the removal of 2,732 hosts.
 
For the following evaluation, we utilized publicly available BGP data from
RouteViews Oregon~\cite{routeviews}, which provides BGP tables every two hours.
As a consequence, we cannot recognize shorter-lived events. This is no inherent
limitation: In productive environments, HEAP can be interfaced with a live
stream of BGP data, e.g.~directly obtained from BGP routers or from services
like BGPmon~\cite{bgpmon}.

\begin{figure}[t!]
 \centering
 \includegraphics[width=0.415\textwidth]{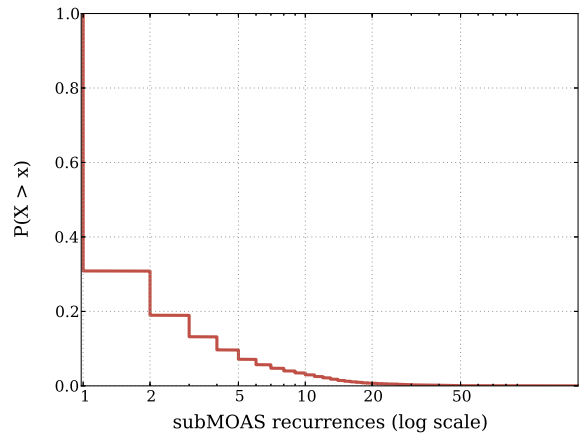}
 \vspace{8pt}
 \caption{Distribution of subMOAS reoccurrences (CCDF).}
 \label{fig:subfreq}
 \vspace{2pt}
\end{figure}

\subsection{Overall Results}\label{subsec:results}

\autoref{fig:subtimes} shows the frequency of subMOAS events observed during
the month of August, 2015. On average, we encountered 88 events every two
hours. The minimum number is 1, the maximum number is 388.
\autoref{fig:subfreq} gives details on subMOAS events that occurred more than
once, i.e.~concerned the same prefixes and ASes. On average, subMOASes recurred
2.2 times, with a maximum of 169 reoccurrences. In the following, multiple
occurences of identical subMOAS events are considered only once.

\begin{table}[t!]
 \footnotesize
 \centering
 \renewcommand{\arraystretch}{1.12}
 \begin{tabularx}{0.45\textwidth}{Xrr}
  & total & in \% \\
  \toprule
  \textbf{All subMOAS events} & \textbf{14,050} & \textbf{100.0\%} \\
  \midrule
  IRR analysis & 5,699 & 40.56\% \\
  topology reasoning & 2,328 & 16.57\% \\
  SSL/TLS scans & 2,639 & 18.78\% \\
  \midrule
  \textbf{Legitimate events (cum.)} & \textbf{7,998} & \textbf{56.93\%} \\
  \bottomrule
 \end{tabularx}
 \vspace{1pt}
 \caption{Overview of HEAP results (combined).}
 \label{table:results_full}
\end{table}

\begin{table*}[t!]
 \footnotesize
 \centering
 \renewcommand{\arraystretch}{1.12}
 \begin{tabularx}{0.9\textwidth}{Xrr|rr|rr|rr|rr}
   \multirow{2}{*}{All subMOAS events: \ \textbf{14,050}} & \multicolumn{2}{c|}{\bf AfriNIC} & \multicolumn{2}{c|}{\bf APNIC} & \multicolumn{2}{c|}{\bf ARIN} & \multicolumn{2}{c|}{\bf LACNIC} & \multicolumn{2}{c}{\bf RIPE} \\
   & total & in \% & total & in \% & total & in \% & total & in \% & total & in \% \\
   \toprule
   \bf Covered subMOAS events & \bf 340 & \bf 2.42\% & \bf 2,020 & \bf 14.38\% & \bf 5,284 & \bf 37.61\% & \bf 574 & \bf 4.09\% & \bf 3,312 & \bf 23.57\% \\
   \midrule\\[-10pt]
   \multicolumn{1}{l}{valid business relationships} & 63 & 0.45\% & 1,042 & 7.42\% & 970 & 6.90\% & n/a & n/a & 1,677 & 11.94\% \\
   \multicolumn{1}{l}{valid resource holdership} & 104 & 0.74\% & 1,298 & 9.24\% & 1,800 & 12.81\% & n/a & n/a & 1,971 & 14.03\% \\
   \midrule\\[-10pt]
   \bf Legitimate events (cum.) & \bf 104 & \bf 0.74\% & \bf 1,397 & \bf 9.94\% & \bf 2,018 & \bf 14.36\% & \bf n/a & \bf n/a & \bf 2,452 & \bf 17.45\% \\
   \bottomrule
 \end{tabularx}
 \vspace{1pt}
 \caption{Overview of HEAP results (IRR filter).}
 \label{table:results_irr}
\end{table*}

During our experiments, we observed a total of 14,050 unique subMOAS events.
Our data sources cover 11,222, i.e.~79.87\% of these events. Hence, our
coverage can still be increased, which suggests that extending HEAP by
additional filters can further improve our legitimization results. By feeding
the subMOAS events into HEAP, we were able to legitimize 56.93\%.
\autoref{table:results_full} presents an overview of individual filter
results. Note that an event might be legitimized by multiple filters: in total,
we obtain 10,666 legitimate events, which amount to 7,998 distinct cases. At
the same time, 5,653 of these cases were legitimized by only a single filter,
i.e.~each filter contributes unique results. The IRR analysis yields 3,660
unique legitimized cases, followed by SSL/TLS scans with 1,244 cases and our
topology reasoning with 749 cases. Overall, we are able to legitimize more than
half of all subMOAS events. We will see later on that HEAP performs
even better under more realistic conditions, i.e.~for alarms relating to
networks that are of high value for an attacker.

\subsection{In-depth Analysis of the IRR Filter} \label{subsec:eval:irr}

With the help of the five IRR databases---provided by AfriNIC, APNIC, ARIN,
LACNIC, and RIPE---we can legitimize 40.56\% of all subMOAS events
observed during our analysis period in August, 2015. We identified 5,971
legitimate causes for 11,530 covered events, i.e.~for cases where the
affected IP prefixes and ASes were registered in one of the IRR databases. Note
that some of these resources are registered in multiple databases. Overall,
we legitimize a total of 5,699 distinct cases out of
10,500 covered unique events.

\autoref{table:results_irr} shows details on the effectiveness of individual IRR
filters at eliminating benign subMOAS events. The highest coverage of events is
provided by ARIN (37.61\%), while AfriNIC and LACNIC cover less than
5\%. At the same time, the ARIN filter legitimizes a comparatively
low fraction of its covered events due to missing maintainer and RPSL
information in the respective IRR data model. In absolute terms, RIPE, ARIN,
and APNIC yield the highest number of legitimized events. LACNIC
removes all privacy-related information from its IRR database snapshots. As a
consequence, none of its covered subMOAS events can be legitimized.

\autoref{table:ripedb} already suggested that filters based on
\textit{org},~\textit{import}, and \textit{maintained\_by}
relations, i.e.~filters utilizing \texttt{ORGANIZATION}, \texttt{ROUTE} and
\texttt{MNTNER} objects, show the potential to perform best due to rich
relations between these resource objects. Our results confirm this assumption.
The most effective filters are based on \texttt{ORGANIZATION} objects in the
ARIN database (11.86\%), followed by \texttt{ROUTE} objects in the
RIPE database (11.17\%). Where applicable, maintainer relations are
highly effective as well (up to 8.31\%). Interestingly, filters that
are based on a combination of \texttt{ORGANIZATION} and \texttt{MNTNER}
objects contribute least to the overall validation results (0.99\% at
most). In total, filter rules that aim at identifying business relationships
can eliminate 25.17\% of all events, while rules that establish
confirmation of resource holdership yield 36.39\% legitimate events. If
we combine them, we find that 40.56\% of all subMOAS events (or 54.28\%
of all covered events) can be legitimized.

\subsection{In-depth Analysis of the SSL/TLS Filter}

\begin{table}[t!]
 \footnotesize
 \centering
 \vspace{-3.5pt}
 \renewcommand{\arraystretch}{1.12}
 \begin{tabularx}{0.45\textwidth}{Xrr}
  & total & in \% \\
  \toprule
  \textbf{SSL/TLS scans} & \textbf{95,486} & \textbf{100.0\%} \\
  \midrule
  same SSL/TLS key & 45,572 & 47.73\% \\
  different SSL/TLS key & 13,202 & 13.83\% \\
  no response (port closed) & 19,119 & 20.02\% \\
  discarded scan results & 17,593 & 18.42\% \\
  \bottomrule
 \end{tabularx}
 \vspace{1pt}
 \caption{Overview of HEAP results (SSL/TLS filter).}
 \label{table:results_tls}
 \vspace{-3pt}
\end{table}

It is worthwile to study the performance of our SSL/TLS filter in more detail.
\autoref{table:results_tls} shows further information about scans to
individual ground truth hosts. In total, we scanned 95,486 SSL/TLS hosts
distributed over 3,236 (23.03\%) subMOAS events. Note that we
discarded 18.42\% of the scan results, for which the subMOAS events changed or
vanished during the scans. Another 20.02\% of our ground truth hosts did not
respond to the validation scans. Overall, 47.73\% of
the retrieved SSL/TLS keys did not change, leading to a total of 2,639
(18.78\%) legitimized subMOAS events.

\begin{figure}[t!]
 \centering
 \includegraphics[width=0.415\textwidth]{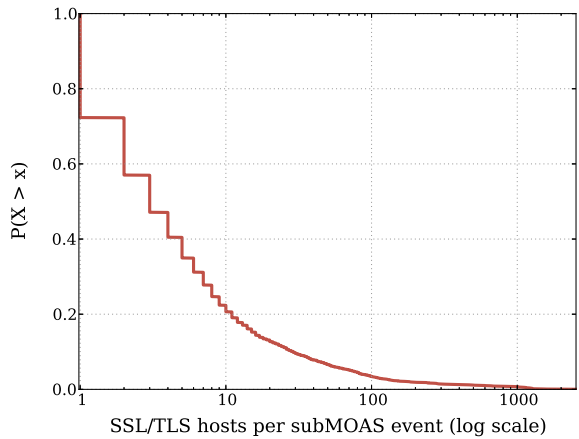}
 \vspace{3pt}
 \caption{Available SSL/TLS hosts per subMOAS event (CCDF).\newline\textit{Events without any such hosts are omitted.}}
 \label{fig:subtlshosts_all}
 \vspace{-4pt}
\end{figure}

Despite the comparatively high number of unusable scan results, we obtain a
relative legitimization rate of 81.55\% for covered subMOAS prefixes,
i.e.~for such prefixes with at least one SSL/TLS-enabled host in our ground
truth data set. To elaborate, \autoref{fig:subtlshosts_all} shows the
distribution of available SSL/TLS hosts per subMOAS prefix. For 47.13\%
of all covered events, our ground truth in fact comprises more than three
available hosts. 20.65\% of these events provide more than ten hosts,
and 3.37\% of them even more than 100. The maximum number of available
hosts is 2,531 with an average of 29.51 hosts per event. These figures actually
allow our SSL/TLS filter to be highly robust against outages of individual
hosts or services, since it is enough for our technique to confirm that
\textit{at least one} cryptographic key on \textit{any} of the affected hosts
inside a prefix remains unchanged during a subMOAS event.
\autoref{fig:subtlsres} further indicates that the fractions of unchanged and
changing keys shift rather slowly over the time frame of one month. Note that
despite the low decline in stable keys, we need to occasionally renew our
ground truth data set nonetheless.

Another interesting fact with respect to the legitimization capabilities of our
SSL/TLS filter is the set of ports, i.e.~network protocols, that contribute to
the validation. \autoref{fig:subtlsports} illustrates that for more than
95\% of covered subMOAS events, respectively events with at least one
SSL/TLS host available, HTTPS servers can be utilized for the validation. Other
protocols like LDAPS, FTPS, XMPPS, and IRCS are apparently ill-suited for our
purposes. While adding robustness against outages of HTTPS services
for half of the HTTPS-validated events (54.96\%), these protocols
contribute as few as 75 (2.83\%) unique legitimate events. These
facts will be taken into account for future re-scans of our ground truth.

\setcounter{figure}{7}
\begin{figure}[t!]
 \centering
 \includegraphics[width=0.435\textwidth]{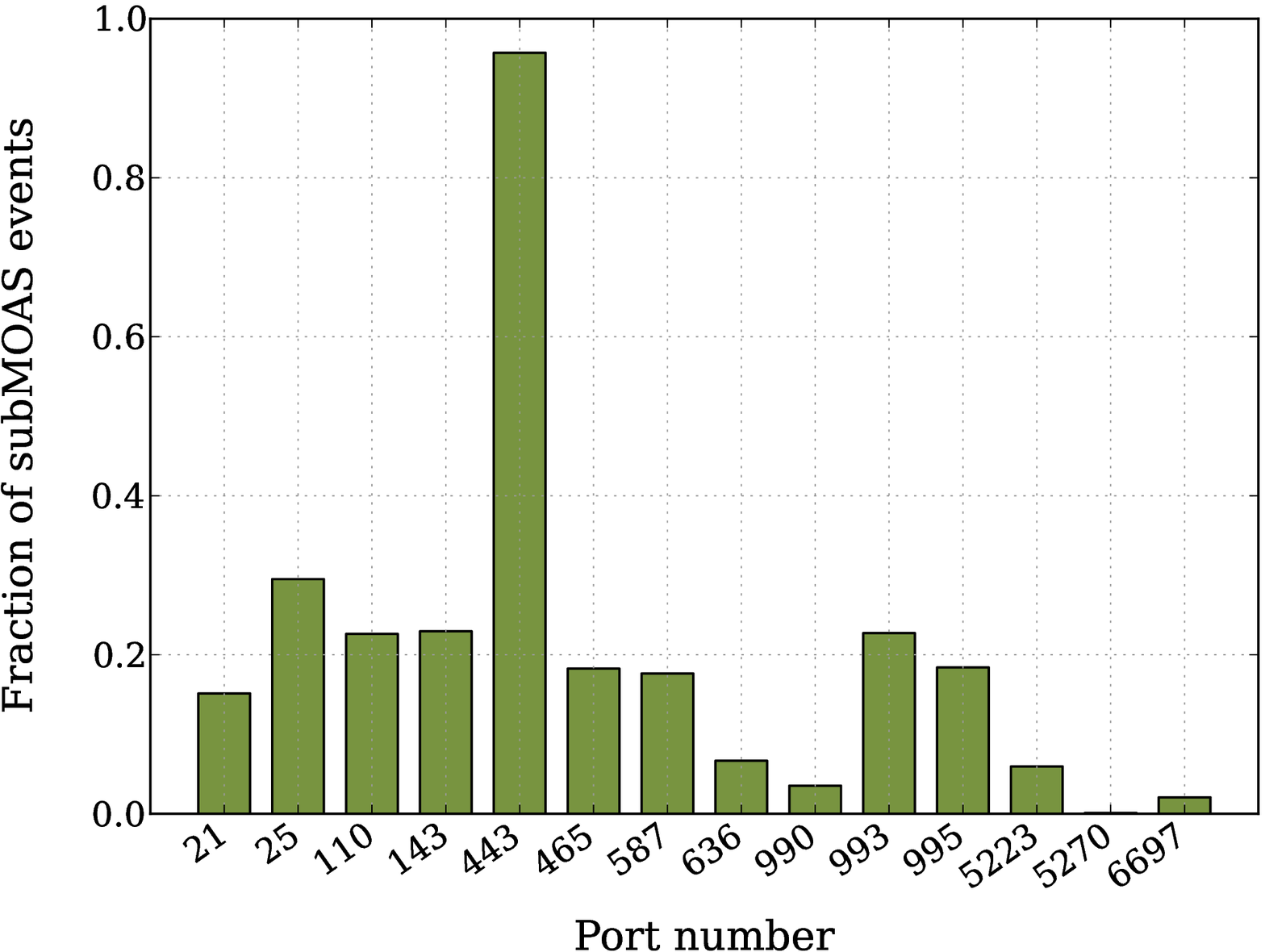}
 \caption{Fraction of available SSL/TLS hosts for covered subMOAS events, broken down by port.}
 \label{fig:subtlsports}
\end{figure}
\subsection{A Practical Case Study} \label{subsec:eval:case}

So far, we evaluated HEAP with respect to its legitimization capabilities of
rather general day-to-day events. To get a more realistic view of its
capabilities to identifiy false positive hijacking alarms in practice, we
conduct a case study as follows. We assume that an attacker has little interest
in hijacking small and insignificant networks, since the corresponding address
space can be easily monitored and, more importantly, has no particular
reputation in terms of globally whitelisted IP ranges. Instead, we assume that
a real attacker would typically hijack smaller parts of large and popular
networks in order to launch and sustain malicious activities. We thus
evaluate HEAP with respect to the more well-known networks. To
this end, we utilize a list of the top one million web sites provided by
\textit{Alexa Inc.}~\cite{alexa}. For each of the domain names in this list, we
perform a reverse DNS lookup. Since multiple (sub)domains can be
hosted on a single server, we obtain a total of 522,655 distinct IP addresses.
We consequently re-assess all subMOAS events during August, 2015 and restrict
the input fed into HEAP to those events that affect the aforementioned
addresses.

\setcounter{figure}{6}
\begin{figure}[t!]
 \centering
 \includegraphics[width=0.425\textwidth]{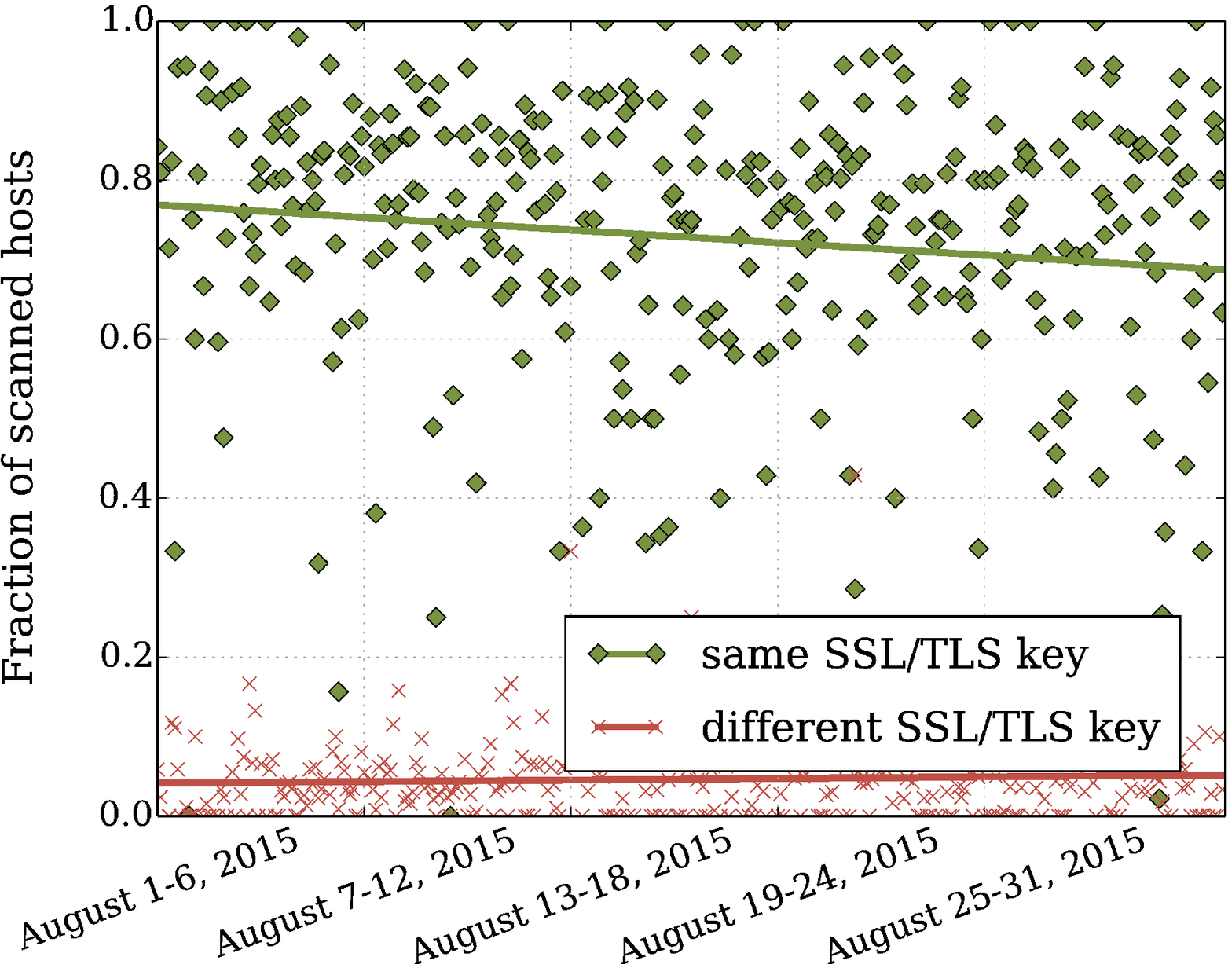}
 \vspace{2.5pt}
 \caption{Fraction of validated SSL/TLS keys during our experiment.}
 \label{fig:subtlsres}
 \vspace{0.5pt}
\end{figure}

\setcounter{figure}{8}
\begin{figure}[t!]
 \centering
 \includegraphics[width=0.415\textwidth]{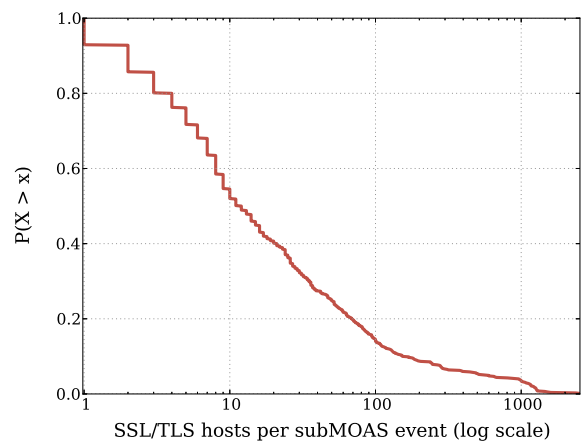}
 \vspace{5.75pt}
 \caption{Available SSL/TLS hosts per subMOAS event (CCDF).\\\textit{Events affecting top 1 million web sites only.}}
 \label{fig:subtlshosts_alexa}
\end{figure}

\begin{table*}[t!]
\centering
\begin{minipage}{0.45\textwidth}
 \footnotesize
 \centering
 \renewcommand{\arraystretch}{1.12}
 \begin{tabularx}{\textwidth}{Xrr}
  & total & in \% \\
  \toprule
  \textbf{All subMOAS events} & \textbf{849} & \textbf{100.0\%} \\
  \midrule
  IRR analysis & 294 & 34.63\% \\
  topology reasoning & 146 & 17.20\% \\
  SSL/TLS scans & 576 & 67.85\% \\
  \midrule
  \textbf{Legitimate events (cum.)} & \textbf{689} & \textbf{81.15\%} \\
  \bottomrule
 \end{tabularx}
 \caption{Overview of HEAP results (combined).\\\textit{Events affecting top 1 million web sites only.}}
 \label{table:results_full:alexa}
 \vspace{-6pt}
\end{minipage}
\quad\quad
\begin{minipage}{0.45\textwidth}
 \vspace{0.7pt}
 \footnotesize
 \centering
 \renewcommand{\arraystretch}{1.12}
 \begin{tabularx}{\textwidth}{Xrr}
  & total & in \% \\
  \toprule
  \textbf{SSL/TLS scans} & \textbf{70,464} & \textbf{100.0\%} \\
  \midrule\\[-7.85pt]
  same SSL/TLS key & 31,888 & 45.25\% \\
  different SSL/TLS key & 6,508 & 9.24\% \\
  no response (port closed) & 17,529 & 24.88\% \\
  discarded scan results & 14,539 & 20.63\% \\[2.85pt]
  \bottomrule
 \end{tabularx}
 \caption{Overview of HEAP results (SSL/TLS filter).\\\textit{Events affecting top 1 million web sites only.}}
 \label{table:results_tls:alexa}
 \vspace{-6pt}
\end{minipage}
\vspace{12pt}
\end{table*}

\begin{table*}[t!]
 \footnotesize
 \centering
 \renewcommand{\arraystretch}{1.12}
 \begin{tabularx}{0.9\textwidth}{Xrr|rr|rr|rr|rr}
  \multirow{2}{*}{All subMOAS events: \ \textbf{849}} & \multicolumn{2}{c|}{\bf AfriNIC} & \multicolumn{2}{c|}{\bf APNIC} & \multicolumn{2}{c|}{\bf ARIN} & \multicolumn{2}{c|}{\bf LACNIC} & \multicolumn{2}{c}{\bf RIPE} \\
  & total & in \% & total & in \% & total & in \% & total & in \% & total & in \% \\
  \toprule
  \bf Covered subMOAS events & \bf 6 & \bf 0.71\% & \bf 217 & \bf 25.56\% & \bf 456 & \bf 53.71\% & \bf 22 & \bf 2.59\% & \bf 139 & \bf 16.37\% \\
  \midrule\\[-10pt]
  \multicolumn{1}{l}{valid business relationships} & 4 & 0.47\% & 62 & 7.30\% & 16 & 1.89\% & 0 & 0.00\% & 57& 6.74\% \\
  \multicolumn{1}{l}{valid resource holdership} & 4 & 0.47\% & 121 & 14.25\% & 59 & 7.00\% & 0 & 0.00\% & 73 & 8.60\% \\
  \midrule\\[-10pt]
  \bf Legitimate events (cum.) & \bf 4 & \bf 0.47\% & \bf 138 & \bf 16.25\% & \bf 62 & \bf 7.30\% & \bf 0 & \bf 0.00\% & \bf 98 & \bf 11.54\% \\
  \bottomrule
 \end{tabularx}
 \vspace{1pt}
 \caption{Overview of HEAP results (IRR filter).\\\textit{Events affecting top 1 million web sites only.}}
 \label{table:results_irr:alexa}
 \vspace{-8pt}
\end{table*}

We find that only a small subset of 849 distinct subMOAS cases out of the full
set of 14,050 events affect these popular networks. At the same time, we see
that the average number of recurring events increases by 13.84\%, which
indicates intentional use of the subMOAS announcements and supports arguments
against misconfiguration or attacks. Hence, one would expect to identify a
larger fraction of legitimate subMOAS events. \autoref{table:results_full:alexa}
present the results. 

We see that HEAP yields a significantly higher legitimization rate of 81.15\%
for subMOAS events that relate to the top one million web sites as compared to
56.93\% for all observed events (see \autoref{table:results_full}. A major
reason for this improvement is an increase in coverage of our methodology. The
combined filter set now covers 98.82\% of the respective events compared to
79.87\% in the section before.  Most notably, the performance of our SSL/TLS
filter is more than three times as high, which is not surprising for
much-frequented networks. As a matter of fact, 73.80\% of all SSL/TLS hosts in
our ground truth data set reside in the popular networks (compare
\autoref{table:results_tls} and \autoref{table:results_tls:alexa}, but
attribute to as few as 6.04\% of all subMOAS events. This finding is also
reflected by \autoref{fig:subtlshosts_alexa}, which shows a significant
difference in the number of available SSL/TLS hosts per subMOAS event as
compared to \autoref{fig:subtlshosts_all}.

The coverage of our IRR filters changes significantly for the
\textit{Alexa}-based events (see \autoref{table:results_irr:alexa}). The
fraction of subMOAS events that affect the ARIN and APNIC service region
increases from 37.61\% to 53.71\%, and from 14.38\% to 25.56\% respectively
(compare to \autoref{table:results_irr}). The remaining IRR filters, in
particular LACNIC and AfriNIC, lose part of their coverage. Altogether, the
overall coverage of subMOAS events increases from 74.73\% to 93.64\%, while the
overall legitimization rate slightly decreases from 40.56\% to 34.63\%.

To put these results into perspective, we use our graph database to identify
the responsible registrars for all of the top 1 million web sites, i.e.~the
respective databases that hold information about corresponding \textit{Alexa}
IP prefixes. Most of these prefixes are registered in the ARIN database
(99.99\%), followed by RIPE (98.86\%) and APNIC (90.77\%). LACNIC (2.38\%) and
AfriNIC (0.54\%) only account for a small number of these web sites. It is
apparent that the largest part of corresponding \texttt{INETNUM} objects is
registered in multiple IRR databases. Such networks often relate to several
regional subsidiaries of worldwide operating companies under independent
administrative control, which possibly explains the slight decrease in our IRR
legitimization rate in spite of an increase in coverage.

\subsection{Feeding Real Alarms into HEAP} \label{subsec:eval:alarms}

With HEAP, we intend to provide a framework that enables reliable assessment of
arbitrary subprefix hijacking alarms. To demonstrate its effectiveness, we
study a set of real alarms reported by BGPmon.net~\cite{bgpstream} during
August, 2015. This set consists of 85 highly suspicious subprefix hijacking
alarms $\hat{\mathcal{A}}$, each given by
\[ \hat{\mathcal{A}} = \{ vp_v \ | \ v \in \Sigma_{AS}, \, p_v \subset \Pi_v \} \, \cup \, \{ ap'_v \ | \ a \in \Sigma_{AS}, \, p'_v \subset p_v \} . \]
\indent During our evaluation, we observed a total of 61 corresponding subMOAS events, for
which we applied our filtering scheme. This lower number of observed events
compared to the full set of reported alarms results from technical aspects of
our experiment design: 7 events lasted for less than two hours, while 9 events
were not classified as strict subMOAS (see \autoref{subsec:setup}).
Another 8 events re-occured and were considered only once. For the remaining
cases, we retroactively applied our IRR and topology-based filters, while we
naturally lack SSL/TLS measurement data from targeted scans.

\autoref{table:bgpstream} shows our overall legitimization results. In total,
our methodology covered 61 (71.76\%) distinct alarms, of which 7 (8.24\%) were
explicitly identified as false positives. Note that BGPmon.net already provides
a highly focused set of alarms, since as few as 85 (0.61\%) out of the total
number of 14,050 subMOAS events were reported during August, 2015. It is thus
highly suprising that these reports still contain nearly 10\% false alarms.
At the same time, these findings evidence the strength of a cross-validation
with HEAP. We plan to provide a public interface to HEAP, which accepts input
alarms in the format as specified above supplemented by timestamps of the
events. Current and future detection systems may then benefit from our
validation scheme as well.

\begin{table}[t!]
 \centering
 \footnotesize
 \renewcommand{\arraystretch}{1.12}
 \setlength{\tabcolsep}{1.75pt}
 \begin{tabularx}{0.4885\textwidth}{Xrr|rr|rr||rr}
  & \multicolumn{2}{c|}{\bf IRR} & \multicolumn{2}{c|}{\bf Topology} & \multicolumn{2}{c||}{\bf SSL/TLS} & \multicolumn{2}{c}{\bf total} \\
  \multirow{2}{*}{reported alarms:~\textbf{85}} & \multicolumn{2}{c|}{\bf analysis} & \multicolumn{2}{c|}{\bf reasoning} & \multicolumn{2}{c||}{\bf scans} & \multicolumn{2}{c}{\bf (cum.)} \\
  & total & in \% & total & in \% & total & in \% & total & in \% \\
  \toprule
  covered alarms & 60 & 70.59\% & 3 & 3.53\% & 1 & 1.18\% & 61 & 71.76\% \\
  \bf false positives & \bf 6 & \bf 7.06\% & \bf 1 & \bf 1.18\% & \bf 0 & \bf 0.00\% & \bf 7 & \bf 8.24\% \\
  \bottomrule
 \end{tabularx}
 \vspace{1pt}
 \caption{HEAP cross-check of BGPmon.net hijacking alarms~\cite{bgpstream}.}
 \label{table:bgpstream}
 \vspace{-8pt}
\end{table}
\subsection{Summary} \label{subsec:summary}

With a thorough evaluation of day-to-day anomalies, we established an
encouraging base line for practical validation of hijacking alarms: we
legitimized 56.93\% of these events. In our case study, we further narrowed
down the search space for practical hijacking attacks by focusing on networks
that host the top one million web sites. Our ability to legitimize 81.15\% of
corresponding events indicates that our methodology performs even better for
such popular networks. These networks may be at higher risk of being attacked
due to their good reputation in whitelists. With an analysis of publicly
reported hijacking incidents, we demonstrated great practical benefits of our
system by identifying nearly 10\% of the alarms as false positives. We are thus
ready to interface our system with that of fellow researchers to receive and
assess their alerts.

Based on our evaluation, we arrived at the following conclusions. First, data
obtained from IRR databases, albeit possibly incomplete, is highly useful to
assess hijacking alarms in practice. Second, our topology reasoning technique
proves to be of equally high effectiveness. Last, but not least, active scans
greatly support a reliable assessment of hijacking alarms. The applicability of
this approach is remarkably high, which, more importantly, relates to a huge
set of SSL/TLS-enabled hosts that remained stable throughout our experiments. We
consequently encourage network operators to ``opt-in'' to HEAP by simply
setting up HTTPS servers with unique SSL/TLS keys in their networks---these
would be automatically found by our ground truth scans and incorporated into
HEAP---ready to be used for validation scans in case of an alarm. In our
evaluation, we further observed striking differences in the deployment of SSL/TLS,
which led to improvements to our plans for regular ground truth scans in the
future.
\section{Conclusion and Outlook} \label{sec:conclusion}

In this paper, we introduced a novel approach to formalize characteristics of
Internet routing. Applied to BGP hijacking, it is suitable to precisely
formulate, classify, and evaluate different kinds of attacks. We utilized this
model to assess impact and traceability of subprefix hijacking attacks. A
concept of general nature, it may serve to establish a basis to address future
research questions on Internet routing. 

Based on our formal attacker model, we derived HEAP, an extensible
filtering system that combines several data sources in order to reliably assess
the validity of hijacking alarms. An automated reasoning technique for given
routing anomalies, it lends itself well to integration with state-of-the-art
and future hijacking detection systems, in particular to cross-check and narrow
down their number of false alarms. In our evaluation, we thoroughly analyzed
the applicability of our approach, and demonstrated its usefulness in practice
by revealing a significant number of false positives in a set of
well-established hijacking reports. We intend to grow our framework into a
public service that makes its data available on a continuous basis. We invite
researchers to feed our system with their conjectural alerts and to further extend the
system by resourceful data sources.

\section*{Acknowledgements}

We thank the reviewers for their constructive comments.

\bibliographystyle{IEEEtran}


\begin{IEEEbiography}[{\includegraphics[width=1in,height=1.25in,clip,keepaspectratio]{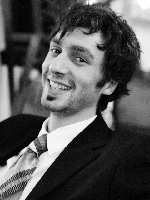}}]{Johann Schlamp}
is a research associate at the Technical University of Munich, Germany. He
studied Computer Science with a minor in Mathematics, and received his
Dipl.-Inf. (M.S.) from the Technical University of Munich in 2009. Since then,
he has been working at the chair of Network Architectures and
Services. In his Ph.D. thesis, he conducted an evaluation of architectural
threats to Internet routing. Johann's research interests lie with interdomain
routing, Internet protocols, and network measurements.
\end{IEEEbiography}

\begin{IEEEbiography}[{\includegraphics[width=1in,height=1.25in,clip,keepaspectratio]{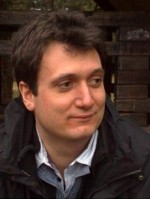}}]{Ralph Holz}
is a lecturer in networks and security at the University of Sydney and a
conjoint researcher at Data61|CSIRO, Australia's premier innovation hub. He
graduated with a Ph.D. from Technical University of Munich, Germany, where he
led the research that first analysed and demonstrated the security shortcomings
of X.509 for the Web. His current research agenda is empirical security, in
particular global-scale analyses of Internet service deployments and
technology. Ralph contributes to the IETF and is an author of two RFCs.
\end{IEEEbiography}

\begin{IEEEbiography}[{\includegraphics[width=1in,height=1.25in,clip,keepaspectratio]{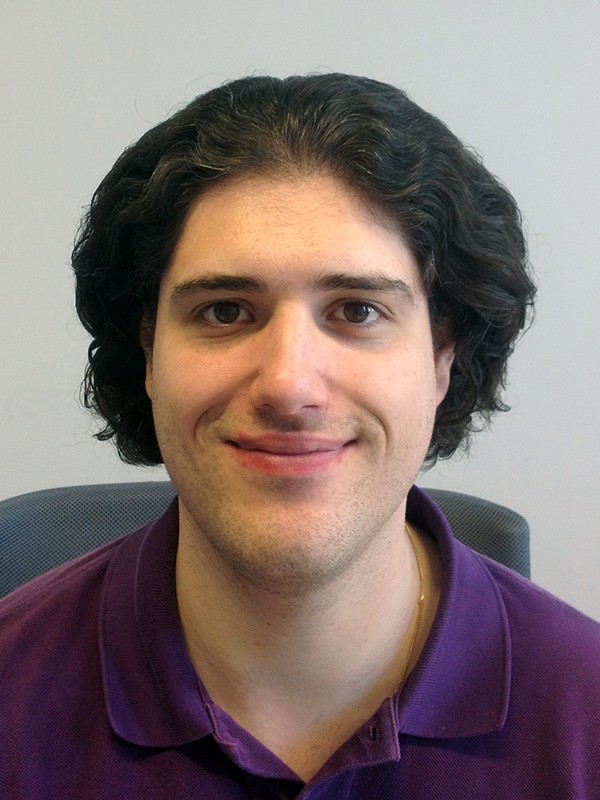}}]{Quentin Jacquemart}
is a postdoctoral research fellow in the SigNet group at the I3S Lab,
University Nice Sophia Antipolis, France. He received his Master's degree in
Computer Science from the University of Liège, Belgium in 2011. Between 2011
and 2015, he was part of the Networking and Security group at Eurécom, Sophia
Antipolis, France, and received his Ph.D. from Telecom ParisTech, France in
2015. His main research interest is the security of computer networks and of
their protocols. His research is funded by the UCN@Sophia Labex.
\end{IEEEbiography}

\vfill
\newpage

\begin{IEEEbiography}[{\includegraphics[width=1in,height=1.25in,clip,keepaspectratio]{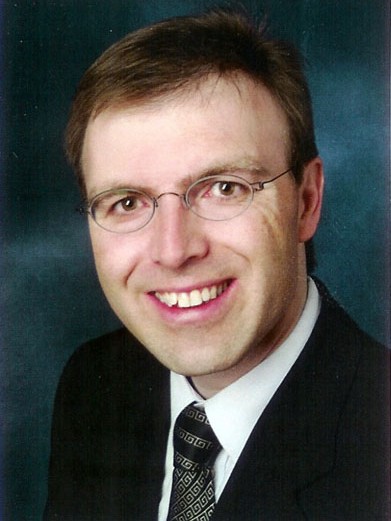}}]{Georg Carle}
is a professor at the Department of Informatics of Technical University of
Munich, Germany, holding the chair of Network Architectures and Services. He
studied Electrical Engineering at the University of Stuttgart, Germany. Studies
abroad included a Master of Science in Digital Systems at Brunel University,
London, and a stay at Ecole Nationale Supérieure des Télécommunications, Paris,
France (now Telecom ParisTech). He received his Ph.D. from the University of
Karlsruhe (now KIT), and worked as a postdoctoral scientist at Eurécom, Sophia
Antipolis, France, and at the Fraunhofer Institute for Open Communication
Systems, Berlin, Germany, prior to being appointed as a professor at University
of Tübingen, Germany.
\end{IEEEbiography}

\begin{IEEEbiography}[{\includegraphics[width=1in,height=1.25in,clip,keepaspectratio]{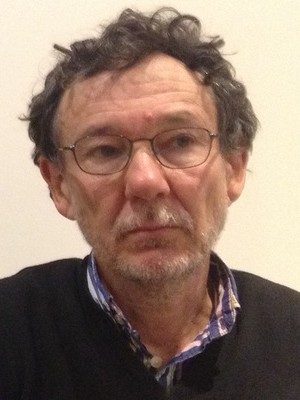}}]{Ernst W. Biersack}
studied Computer Science at the Technical University of Munich, Germany and at
the University of North Carolina at Chapel Hill, US. He received his Dipl.
Inform. (M.S.) and Dr. rer. nat. (Ph.D.) degrees in Computer Science from the
Technical University of Munich, and his Habilitation à Diriger des Recherches
from the University of Nice, France. From March 1989 to February 1992 he was a
Member of Technical Staff with the Computer Communications Research Group of
Bell Communications Research, Morristown, US. From March 1992 to mid 2014 he
was a Professor at Eurécom, Sophia Antipolis, France.
\end{IEEEbiography}

\vfill

\end{document}